\documentclass[letterpaper, 10pt, conference]{ieeeconf}      

\usepackage{xcolor}
\usepackage{amsmath,amssymb,amsfonts}
\usepackage{mathtools}
\usepackage{graphicx}

\IEEEoverridecommandlockouts       
\usepackage[ruled,vlined]{algorithm2e}
\newtheorem{theorem}{Theorem}
\newtheorem{lemma}{Lemma}
\title{\LARGE \bf
$\mathcal{H}_2/\mathcal{H}_{-}$ Distributed Fault Detection and Isolation for Heterogeneous Multi-Agent Systems
}

\author{Thiem V. Pham$^{1}$ and Quynh T.~T. Nguyen$^{1, \star}$
\thanks{$^{1}$Thiem V. Pham and Quynh T.~T. Nguyen  are with Thai Nguyen University of Technology (TNUT), Viet Nam,
        {\tt\small{phuthiem @tnut.edu.vn}, \tt\small{nttquynh-dldk@tnut.edu.vn}}}}%

\begin{document}

\maketitle
\thispagestyle{empty}
\pagestyle{empty}

\begin{abstract}
The paper deals with the problem of distributed fault detection and isolation (FDI) for a group of heterogeneous multi-agent systems. The developed formation for the FDI is taken into account as a distributed observer design methodology, where the interaction between the agent and its neighbors is described as a vector of distributed relative output measurements. Based on two performance indexes $\mathcal{H}_2$ and $\mathcal{H}_{-}$, sufficient conditions are given to ensure the residual signals robust to the disturbances and sensitive with respect to the fault signals. In addition, we show that by using our proposed approach, each agent is able to estimate both its own states and states of its nearest neighbors in the presence of disturbances and faults.  Finally, numerical simulations are provided to demonstrate the effectiveness of the theoretically analyzed results.
\end{abstract}
\section{Introduction}
Multi-agent systems (MASs) have received considerable attention in recent years, and they have been applied in a wide variety of areas, such as automotive control system \cite{Pham2019b}, unmanned aerial vehicles \cite{Pham2020c}, sensor network \cite{Pham2020d}, etc. Besides, MASs are vulnerable to faults and attacks from the network.  The faults or the attack occurring at an agent might spread to other agents.  It means that the  neighbor's agent should be affected by these faults or attacks through a communication topology of MASs \cite{Pham2020b}, \cite{Nguyen2020a}. The imperative purpose of fault detection and isolation (FDI) of MASs is thus becoming a very crucial, more challenging.

In fault detection and isolation problems for MASs, there are three methodologies: centralized, decentralized and distributed filter design. Among three aforementioned approaches, the centralized architecture is the fewest attraction because the structure of the MASs is distributed and not all measurements are available to each agent. Moreover, MASs could be burdened on a complex computation cost when the number of agents increases. Concerning the decentralize approach, a sufficient condition based on the $\mathcal H_{\infty}$ performance for large-scale systems is obtained by applying Lyapunov stability theory \cite{Li2009a}. Next, a local/decentralized detection and isolation for multi-robot systems was proposed \cite{Arrichiello2015}, where each robot is able to detect and isolate faults occurring on other robots. However, the decentralized approach is constructed by an observer, which contains the model of entire system. Thus, compared to the centralized and decentralized FDI methods, the distributed FDI for MASs has been the most attractive topic, because of using fewer network resources and having lower computation complexity. In the distributed FDI methodology, there are two attractive trends in considering, the first trend introduce a consensus protocol, which establishes a distributed dynamic model of the MASs for specific agent \cite{Shames2011, Liu2016, Quan2016, Gao2017}.  Each agent in the network can update its information according to the consensus protocol and can send and receive information with its neighbor agents. Then, by constructing a bank of observers, each agent can detect not only its faults but also faults of its own neighbor agents. However, as the number of agents increases in the network, the FDI will place a heavy computational burden on the entire system. The second trend uses the sensor measurements both locally and from the agent's neighbors \cite{Davoodi2016, Nguyen2017d}.

On the basis of the above review, there are some limitations on the distributed FDI for MASs. Firstly, most of the literature considers the FDI for network of \textit{homogeneous dynamics} rather than \textit{heterogeneous dynamics}, which are different dynamics, as well as sensor faults are not considered in \cite{Liu2016}, \cite{Quan2016}, \cite{Shames2011}. Secondly, the FDI will suffer from a heavy computational burden on the whole system. To overcome this problem, we use the relative output measurements such as \cite{Nguyen2020a} to construct the virtual model of each agent. Thirdly, in order to detect faults, unknown input observer is a powerful technique, where the perfect unknown input decoupling condition is needed to guarantee. If this condition is not satisfied, which is common in practice, the methodologies proposed in \cite{Gao2017}, \cite{Quan2016} could not be applied. Therefore, instead of employing unknown input observer with the perfect unknown input decoupling condition, the distributed observer based on Luenberger is proposed in this paper, which has a simple structure.  Moreover, the residual generation problem can be formulated as an $\mathcal{H}_2$ optimal filtering problem (the Kalman filtering) \cite{Ding}. Finally, although most of the proposed methodologies in previous works can achieve fault detection and isolation, there are rare methodologies, which achieves FDI and state estimation objectives at the same time.

\textbf{Contributions}. Motivated by the above works, the problem of distributed FDI for a network of heterogeneous MASs is addressed to detect and isolate the faults.  A relative model is constructed by utilizing a combination of local sensor information of the nearest neighboring agents (called a vector of distributed relative output measurements). Therefore, the main contributions of this paper are summarized as follows: First, we develop a distributed observer for a team of time-invariant MASs, which utilizes a vector of distributed relative output measurements. To guarantee the existence of at least an observer matrix gain, the detectable property of closed loop system should be guaranteed. Furthermore, by using the distributed Luenberger observer, each agent is able to estimate its own states and the states of its nearest neighbors in the presence of the disturbances, faults, and the control inputs. Secondly, a multi-objective optimization, $\mathcal{H}_2$ and $\mathcal{H}_{-}$ performance indexes, is introduced to robustness against the disturbance signal on the residual signals and the sensitivity for the fault signals. Similar to the work in \cite{Nguyen2017d} the formulation proposed in this paper, in which each agent can detect not only its own faults but also the faults of its neighbors. Then, the distributes fault isolation strategy is proposed corresponding to the residual signals of the network.

This paper is organized as follows. In Section II, we describe the system description and problem formulation. The LMI-based solution to the distributed FDI problem is developed in Section III. To demonstrate the validity of the proposed approach, a numerical example is given in Section IV which is followed by conclusions in Section V.

\textbf{Notations}. The notation used in this paper is fairly standard. For a given matrix $A, A^T$ and $Trace(A)$ denote its transpose and trace, respectively. Using the notation $G:=(A,B,C,D)$
\begin{align*}
G:=\left[
\begin{array}{c|c}
  A & B \\ \hline
  C & D \\
 \end{array}
\right] , [G_1\;\;G_2] :=\left[
\begin{array}{c|c}
  A & B_1 \;\;\;\;B_2 \\ \hline
  C & D_1 \;\;\;\;D_2 \\
 \end{array}
\right] 
\end{align*}
where $G_1:=(A,B_1,C,D_1)$ and $G_2:=(A,B_2,C,D_2)$. We drop the argument "s" in transfer matrices. The transfer matrix $G$ is proper if $G(\infty)=D$ and $G$ is strictly proper if $G(\infty)=0$. $\mathbb{RH}_{\infty}$ and $\mathbb{RH}_{2}$ denotes the set of stable and strictly proper transfer matrices. The $\mathcal H_2$ norm of $G$ is calculated as $\|G\|_2^2=Trace(B^TYB)=Trace(CQC^T)$ where $Y$ and $Q$ represent respectively the observability and controllability Gramaian ($A^TY+YA+C^TC=0$ and $AQ+QA^T+BB^T=0$). The notation $\|.\|_p$ denotes the $\mathcal{H}_2$ or $\mathcal{H}_{-}$ norm. Finally, we use $*$ to denote the symmetry entries of symmetry matrices. 
\section{Problem Formulation}
\subsection{System description}
Consider a network of $N$ heterogeneous agents, where each agent is expressed by a linear dynamic model such as
\begin{align}\label{1}
\begin{aligned}
\dot x_i(t)&= {A}_{i}x_i(t)+{B}_{i}u_i(t)+{B}_{fi}f_i(t)+{B}_{di}d_i(t)\\
y_i(t)&= {C}_{i}x_i(t)+{D}_{fi}f_i(t)+{D}_{di}d_i(t)
\end{aligned}
\end{align}
where $x_i(t) \in \mathbb{R}^{n_i}$ denotes the sate vector, $f_i(t)\in \mathbb{R}^{n_{fi}}$ denotes the faults signal (sensor faults, actuator faults and process faults), $d_i(t) \in \mathbb{R}^{n_{di}}$ denotes the disturbance and noise signals or uncertain components, $u_i(t) \in \mathbb{R}^{n_{u_i}}$ denotes the control inputs, and $y_i(t) \in \mathbb{R}^{n_{y_i}}$ denotes the measured outputs for the agent $i^{th}$ with $n_{y_i} \ge n_{f_i}$. In addition, system matrices ${A}_{i}, {B}_{i}, {B}_{fi}, {B}_{di}, C_{i}, {D}_{fi}$ and ${D}_{di}$  are constant matrices. The fault matrices ${B}_{fi}$ and ${D}_{fi}$ are specified according to faults that are to be detected in the components, sensors, and actuators. In the rest of paper, we omit the term $"t"$ in $x_i,u_i,f_i,d_i, y_i$. 

With the assumption that all agents have the same number of outputs, we assume that each agent not only measures the trivial absolute output signal $y_i$ but also is equipped with the sensors for the relative output measurements, that is $z_{ij} = y_i - y_j, j\in N_i = {i_1,i_2,...,i_{|Ni|}} \subseteq [1,N]$ denotes the set of agents that agent $i^{th}$ can sense ($i^{th}$ agent's neighbors). The communication topology among the $N$ agents is represented by an undirected graph $\mathcal G = (V,\mathcal E)$ \cite{Pham2019c}, consisting of the node set $V = \{1,2,...,N\}$ and the edge set $\mathcal E \in V \times V$. We are interest in, at each time instant, the information available to agent $i^{th}$, which is the relative measurement of other agents with respect to itself. 

The vector relative output of agent $i^{th}$ is thus expressed \begin{align*}
    z_i&=(z_{i_{i_1}},z_{i_{i_2}},\cdots,z_{i_{|Ni|}})^T\\
    &=(y_i - y_{i_1},y_i - y_{i_2},\cdots, y_i - y_{i_{|Ni|}})^T
\end{align*}
The relative model for $i$ agent could be expressed
\begin{align}\label{2}
\dot{x}_{Ni} &= \mathcal{A}_i x_{Ni}+\mathcal{B}_{ui} u_{Ni}+\mathcal{B}_{fi} f_{Ni}+\mathcal{B}_{di} d_{Ni}\nonumber\\
z_i&= \bar{\mathcal C}_i x_{Ni}+\bar{\mathcal D}_{fi}f_{Ni}+\bar{\mathcal D}_{di}d_{Ni}
\end{align}
where $x_{Ni}=(x_i^T,\cdots,x_{i_{|Ni|}}^T)^T, u_{Ni}=(u_i^T,\cdots,u_{i_{|Ni|}}^T)^T$ denote a state and control input vectors. $d_{Ni}=(d_i^T,\cdots,d_{i_{|Ni|}}^T)^T, f_{Ni}=(f_i^T,\cdots,f_{i_{|Ni|}}^T)^T$ stands for a disturbance and fault vectors. $\mathcal{A}_i, \mathcal{B}_{ui}, \mathcal{B}_{fi}, \mathcal{B}_{di}$ and $ \bar{\mathcal C}_i, \bar{\mathcal D}_{fi}, \bar{\mathcal D}_{di}$ are defined with the following notation.

For given matrices $H_i^{n_i \times n_i}, H_{i1}^{n_{i1} \times n_{i1}},...,H_{i_{|Ni|}}^{n_{i_{|Ni|}} \times n_{i_{|Ni|}}}$, where $\mathcal H_i$ denote the $\mu_i \times \xi_i$ matrices
\begin{align*}
    \mathcal H_i &= \left[ \begin{matrix}
    {{{H}_{i}}}&0&0&...&0\\
    0& {{{H}_{i_1}}}&0&...&0\\
    \vdots &\vdots&\vdots&\ddots&0\\
    0&0&0&...&{{{H}_{i_{|Ni|}}}}
    \end{matrix} \right]\\
    \bar{\mathcal H}_i &= \left[ \begin{matrix}
    {{{H}_{i}}}&{{-{H}_{i_1}}}&0&...&0\\
    \vdots &\vdots&\vdots&\ddots&0\\
    {{{H}_{i}}}&0&0&...&{{-{H}_{i_{|Ni|}}}}
    \end{matrix} \right]
\end{align*}
where $\mu_i = n_i +\sum_{j=1}^{|Ni|}n_{ij}$ and $\xi_i =\sum_{j=1}^{|Ni|}m_{ij}$. $\mathcal{A}_i \in \mathbb{R}^{\mu_i \times \mu_i}, \mathcal{B}_{ui} \in \mathbb{R}^{\mu_i \times \mu_{ui}}, \mathcal{B}_{di} \in \mathbb{R}^{\mu_i \times \mu_{di}}$ and $\mathcal{B}_{fi} \in \mathbb{R}^{\mu_i \times \mu_{fi}}$ where $\mu_i = n_i +\sum_{j=1}^{|Ni|}n_{ij}, \mu_{ui} = n_{ui} +\sum_{j=1}^{|Ni|}n_{u_{ij}}, \mu_{fi} = n_{fi} +\sum_{j=1}^{|Ni|}n_{f_{ij}}$ and $\mu_{di} = n_{di} +\sum_{j=1}^{|Ni|}n_{d_{ij}}$. The matrices $\bar{\mathcal C_i} \in \mathbb{R}^{\xi_{yi} \times \mu_i}, \bar{\mathcal D}_{di} \in \mathbb{R}^{\xi_{yi} \times \xi_{di}}$ and $\bar{\mathcal D}_{fi} \in \mathbb{R}^{\mu_{yi} \times \xi_{fi}}$ where $\xi_i =\sum_{j=1}^{|Ni|}m_{ij}, \xi_{yi} =\sum_{j=1}^{|Ni|}m_{y_{ij}}, \xi_{di} =\sum_{j=1}^{|Ni|} m_{d_{ij}}$ and $\xi_{fi} =\sum_{j=1}^{|Ni|} m_{f_{ij}}$.

In this paper, we use the following version of the bounded real lemma.
\begin{lemma}\cite{Ding}\label{l1}
	Given system $ C(sI- A+{LC})^{-1}( B_d-{LD}_d)$ and suppose that $( C,  A)$ is detectable, $ D_d$ has full row rank with $ D_d  D_d^T=I$ and $rank \begin{bmatrix}
			{{{ A- j \omega I}}}&{{ B_d}}\\
			{{ C}}&{{ D_d}}
	\end{bmatrix}$ has full rank for all $\omega \in [0, \infty ]$, then the minimum 
	\begin{align*}
		\mathop {min}\limits_{ L}\left \| { C} (sI-{{ A}}+{ L} { C})^{-1}({{ B}}_{d}-{ L} { D}_{d}) \right \|_2 =trace ({ C}  Y { C}^T)^{\frac {1} {2}}
	\end{align*}
	is achieved by 
	$  L =  Y  C^T +{ B}_{d}  { D}_{d}^T$
	and matrix $ Y \ge 0$ solves the Riccati equation
	\begin{align*}
			&({ A}-{ B}_{d} { D}_{d}^T { C}) Y + Y({ A}-{ B}_{d} { D}_{d}^T { C})^T -  Y  { C}^T { C}  Y \nonumber\\
			&+{ B}_{d} { B}_{d}^T -{ B}_{d}  D^T_{d}{ D}_{d} { B}_{d}^T =0
	\end{align*}
\end{lemma}
\begin{lemma}\cite{Guang-RenDuan2013} \label{l2}
	We take into account $G(s):=( A, B, C, D)$, $ A$ is stable. 
	$G(s) \in \mathcal{RH}_{\infty}^{m \times k}$ being injective $\forall \omega$
	\begin{align*}
	    rank \begin{bmatrix}
			{{{ A- j \omega I}}}&{{ B}}\\
			{{ C}}&{{ D}}\\
	\end{bmatrix}=n+k; {DD}^T-\gamma^2 I >0
	\end{align*}
	Then $\left \| G(s) \right \|_{-} > \gamma$ if only if $ X= X^T$ such that
	\begin{align*}
		&XA+ A^TX+ C^T C\\
		&+(XB+ C^T D)(\gamma^2 I- D^T D)^{-1}( B^T X+ D^T C)>0
	\end{align*}
\end{lemma}
\subsection{Distributed Observer}
A distributed observer for the $i$ agent, based on the relative model, is shown as
\begin{align}\label{3}
\begin{aligned}
\dot{\hat{x}}_{Ni} &= \mathcal{A}_i \hat x_{Ni}+\mathcal{B}_{ui}u_{Ni}+{L}_i(z_i-\hat z_i)\\
\hat z_i&= \bar{\mathcal C}_i\hat x_{Ni}
\end{aligned}
\end{align}
and the residual generator in distributed FDI system should described  the inconsistency between the actual system variables and the mathematical model. It could be realized as a composition of state observer, and responses to faults, disturbances and modeling error
\begin{align}\label{3a}
\begin{aligned}
r_i&=z_i-\hat z_i\\
&=\bar{\mathcal C}_i e_{Ni}+\bar{\mathcal D}_{fi}f_{Ni}+\bar{\mathcal D}_{di}d_{Ni}
\end{aligned}
\end{align}
where $\hat x_{Ni} \in \mathbb{R}^{\mu_i}$ denote the state estimation derived by the observer. $L_i \in \mathbb{R}^{\mu_i \times \xi_{i} }$ is the observer  gain matrices,  and is to be determined. 

We define a state estimation error between observer (\ref{3}) and the relative model (\ref{2})  as $e_{Ni}=x_{Ni}-\hat x_{Ni}$. It follows that the dynamic of $e_{Ni}$ can be re-expressed as
\begin{align}\label{4}
\dot e_{Ni} &=(\mathcal{A}_i-{L}_i \bar{\mathcal C}_i) e_{Ni} +(\mathcal{B}_{di}-{L}_i \bar{\mathcal D}_{di}) d_{Ni}\notag\\
&+(\mathcal{B}_{fi}-{L}_i \bar{\mathcal D}_{fi}) f_{Ni}\\
r_i&=\bar{\mathcal C}_i e_{Ni}+\bar{\mathcal D}_{fi}f_{Ni}+\bar{\mathcal D}_{di}d_{Ni})\label{4b}
\end{align}
By taking the Laplace transforms, it is easy to show that
\begin{align}\label{4a}
r_i=T_{r_{i}d_{Ni}}d_{Ni}+T_{r_{i}f_{Ni}}f_{Ni}
\end{align}
where 
\begin{align*}
&[T_{r_{i}d_{Ni}}T_{r_{i}f_{Ni}}]\\
&=
\left[\begin{array}{c|c}
  \mathcal{A}_i-{L}_i \bar{\mathcal C}_i & \mathcal{B}_{di}-{L}_i \bar{\mathcal D}_{di} \mathcal{B}_{fi}-{L}_i \bar{\mathcal D}_{fi} \\ \hline
  \bar{\mathcal C}_i & \bar{\mathcal D}_{di}\;\;\;\;\;\;\;\;\;\;\;\;\bar{\mathcal D}_{fi} \\
 \end{array}\right]
\end{align*}
are the transfer matrices from disturbances and faults to residual, respectively. 

The proposed distributed FDI problem is now to answer the question \textit{"How each agent can detect and isolate not only both its own faults and simultaneously estimate states, but also faults and states of its neighbors using the relative outputs $z_i - z_{i_{i1}},...,z_i - z_{i_{Ni}}$"}. Moreover, it is easy from (\ref{4a}) that if we ignore the faults $f_{Ni}$, the residual $r_i$ only depends on the disturbances $d_{Ni}$ and the optimization problem such as the $\mathcal H_2$ filtering problem (the Kalman filtering). Similarly, if we disregard $d_{Ni}$ then the optimization problem is the $\mathcal H_{-}$ filtering problem. 

We would like to  solve the following multi-criterion optimization problem
\begin{align*}
I.& \;\left \| T_{r_i d_{Ni}}  \right \|_{2} < \gamma_1\\
II.& \;\left \| T_{r_i f_{Ni}}  \right \|_{-} > \gamma_2\\
III.& \;(\mathcal{A}_i-{L}_i \bar{\mathcal C}_i) \;\text{is \;stable}
 \end{align*}
The first and second constraints ensure a trade-off between the robustness against the disturbances $d_{Ni}$ and the sensitivity for the faults $f_{Ni}$. The third condition ensures that the closed loop system (the relative model (\ref{2}) and distributed observer (\ref{3})) is stable and ensures dynamic errors of state estimation converges. To do this, we need to find $L_i$ such that satisfies multi-criterion optimization problem from I to III. We propose $L_i = \textbf L_i+\Delta L_i$, where $\textbf L_i$ is solution of Riccati equation in \textit{Lemma \ref{l1}}, and instead of finding $L_i$, we now need to find $\Delta L_i$ such as 
\begin{align}\label{a}
&\;\;\;\;\;\;\;\;\;\;\;\;\;\mathop {maximize} \beta_2 \gamma_2-\beta_1\gamma_1\\
&subject \;\;to\notag\\
&\hspace*{1cm}I. \;\left \| T_{r_i d_{Ni}}  \right \|_{2} < \gamma_1\notag\\
&\hspace*{1cm}II. \;\left \| T_{r_i f_{Ni}}  \right \|_{-} > \gamma_2\notag\\
&\hspace*{1cm}III. \;[\mathcal{A}_i-(\textbf L_i+\Delta L_i) \bar{\mathcal C_i}] \;\text{is \;stable}\notag
 \end{align}
We can see that the fault detection and robust against to disturbance can be improved by maximizing $\gamma_2$ and minimizing $\gamma_1$. Moreover, $\beta_1, \beta_2$ depends on designer, who want to emphasis the role of fault detection or of robust against the disturbance.

\section{Main Results}
There are three performance indices from I to III  that must be satisfied simultaneously for solving the distributed FDI problem. We assume that the pair ($\mathcal{A}_i, \bar{\mathcal C}_i$) is detectable. The underlying idea adopted here for solving such optimization problems is that the multi-objective optimization problem can be reduced to a single optimization problem with constraints. To this end, the well-established robust control theory and LMI-techniques have been used. In the following theorem, a feasible solution to the distributed FDI problem is obtained by simultaneously considering these indices.
\begin{theorem}\label{thm1}
Consider the relative model system (\ref{2}) and distributed observer (\ref{3}). The closed-loop system will be stable and simultaneously satisfied problems I -- III,  if there exits positive given scalar $\beta_1>0, \beta_2>0$ and positive matrices $N_i,P_i$ for the following optimization $\mathcal H_2/ \mathcal H_{-}$ problem
\begin{align}\label{5f}
\mathop{maximize}\limits_{N_i,P_i} \beta_2 \gamma_2-\beta_1 \gamma_1
\end{align}
subject to
\begin{align}
\begin{bmatrix}\label{10}
{{Q_{i}}}&{{N_i^T}}\\
{{*}}&{{P_{i}}} 
\end{bmatrix} \geq  &0\\
trace(Q_i) - \gamma_1^2 +trace (\mathcal{C}_i Y_i\bar{\mathcal C}_i^T)\leq&  \\
\mathbf{A}_i P_i - N_i\bar{\mathcal C}_i +P_i\mathbf{A}_i^T  -\bar{\mathcal C}_i^TN_i^T+ \bar{\mathcal C}^T \bar{\mathcal C}_i <& 0 \\
\begin{bmatrix}\label{5b}
\bar{\mathcal D}_{fi}^T \bar{\mathcal D}_{fi}-\gamma_2^2 I&{{{P}_i\mathbf{B}_{fi}-N_i\bar{\mathcal D}_{fi}+\bar{\mathcal C}_i^T \bar{\mathcal D}_{fi}}}\\
{{*}}&{{F_i}}
\end{bmatrix} >& 0
\end{align}
where $F_i=\mathbf{A}_iP_i - N_i\bar{\mathcal C}_i +P_i\mathbf{A}_i^T  -\bar{\mathcal C}_i^TN_i^T+ \bar{\mathcal C}_i^T \bar{\mathcal C}_i, \mathbf{A}_i=\mathcal{A}_i - \textbf L_i \bar{\mathcal {C}}_i, \mathbf{B}_{fi}=\mathcal{B}_{fi} - \textbf L_i \bar{\mathcal {C}}_i$. The observer gain matrix  is calculated as
\begin{align*}
    L_i =\textbf L_i+P^{-1}_iN_i
\end{align*}
where $\textbf L_i=Y_i\bar{\mathcal C}_i^T+\mathcal{B}_{di} \bar{\mathcal D}_{di}^T$ and $\Delta L_i = P^{-1}_i N_i$.
\end{theorem}

\begin{proof}
There are two parts in our proof. The norm of matrix transfer function of disturbance will be constructed in step 1, where the performances \textit{I} and \textit{III}  corresponding to equation (\ref{a}) will be proved in this step. In step 2, the $\mathcal H_{-}$ use to compute the norm of matrix transfer of faults, where the equation (\ref{5b}) is derived from the performance \textit{II}.

In the first step, the dynamic of error between observer (\ref{3}) and the relative model (\ref{2}) is expressed
\begin{align}\label{5}
\dot e_{Ni} (t)&=(\mathcal{A}_i-\textbf L_i \bar{\mathcal C}_{i}) e_{Ni}(t) +(\mathcal{B}_{di}-\textbf L_i \bar{\mathcal D}_{di}) d_{Ni}(t) \notag\\
&+(\mathcal{B}_{fi}-\textbf L_i \bar{\mathcal D}_{fi}) f_{Ni}(t)+\varrho_{Ni}(t)
\end{align}
where
\begin{align*}
    \varrho_{Ni}(t)&=-\Delta L_i (\bar{\mathcal C}_{i} e_{Ni}(t) +\bar{\mathcal D}_{di} d_{Ni}(t) + \bar{\mathcal D}_{fi} f_{Ni}(t))\\
    L_i &= \textbf L_i+\Delta L_i
\end{align*}

For $f_{Ni}(t) =0$, by taking the Laplace transform of (\ref{5}) and (\ref{4b}), we have
\begin{align*}
e_{Ni}&=(sI-\mathcal{A}_i+\textbf L_i \bar{\mathcal C}_{i})^{-1} [(\mathcal{B}_{di}-\textbf L_i \bar{\mathcal D}_{di}) d_{Ni} +\varrho_{Ni}]\\
r_i&=[\bar{\mathcal C}_{i}(sI-\mathcal{A}_i+\textbf L_i \bar{\mathcal C}_{i})^{-1} [(\mathcal{B}_{di}-\textbf L_i \bar{\mathcal D}_{di}) d_{Ni}\notag\\ 
&+\varrho_{Ni}]+\bar{\mathcal D}_{di}d_{Ni}]
\end{align*}
Let $T_{r_{i}d_{Ni}}$ denote the dynamic part of transfer matrix from $d_{Ni}$ to $r_i$
\begin{align}\label{6}
T_{r_{i}d_{Ni}}*d_{Ni}&=\bar{\mathcal C}_{i} (sI-\mathcal{A}_i+\textbf{L}_i \bar{\mathcal C}_{i})^{-1}[((\mathcal{B}_{di}-\textbf{L}_i \bar{\mathcal D}_{di})\nonumber\\
&+\bar{\mathcal D}_{di})d_{Ni} +\varrho_{Ni}]
\end{align}
The dynamic of output $\varrho_{Ni}(t)=-\Delta L_i (\bar{\mathcal C}_{i} e_{Ni}(t) +\bar{\mathcal D}_{di} d_{Ni}(t) + \bar{\mathcal D}_{fi} f_{Ni}(t))$, for $f_{Ni}(t) =0$, is represented in Laplace domain where $e_{Ni}$ is rebuilt from (\ref{5}) such that
\begin{align*}
\varrho_{Ni}&=-\Delta L_i (\bar{\mathcal C}_{i} e_{Ni} +\bar{\mathcal D}_{di} d_{Ni})\\
&=-\Delta L_i(\bar{\mathcal C}_{i} (sI-\mathcal{A}_i+\textbf L_i \bar{\mathcal C}_{i})^{-1}[(\mathcal{B}_{di}-{L}_i \bar{\mathcal D}_{di})d_{Ni}\\
&+\varrho_{Ni})+\bar{\mathcal D}_{di} d_{Ni}]\\
&=(I +\Delta L_i(\bar{\mathcal C}_{i} (sI-\mathcal{A}_i+\textbf L_i \bar{\mathcal C}_{i})^{-1})^{-1}(-\Delta L_i)\\
&\times (\bar{\mathcal C}_{i} (sI-\mathcal{A}_i+\textbf L_i \bar{\mathcal C}_{i})^{-1}(\mathcal{B}_{di}-\textbf L_i \bar{\mathcal D}_{di})+\bar{\mathcal D}_{di})d_{Ni}
\end{align*}
Subtitled $\varrho_{Ni}$ into (\ref{6}), after some algebra the transfer matrix of disturbance can be reformulated
\begin{align}\label{7}
T_{r_{i}d_{Ni}}
&=\bar{\mathcal C}_{i} (sI-\mathcal{A}_i+\textbf L_i \bar{\mathcal C}_{i})^{-1}(\mathcal{B}_{di}-\textbf L_i \bar{\mathcal D}_{di})+\bar{\mathcal D}_{di}\notag\\
&+\bar{\mathcal C}_{i}(sI-\mathcal{A}_i+\textbf L_i \bar{\mathcal C}_{i}+\Delta L_i \bar{\mathcal C}_{i})^{-1}(-\Delta L_i)\\
&\times(\bar{\mathcal C}_{i} (sI-\mathcal{A}_i+\textbf L_i \bar{\mathcal C}_{i})^{-1}(\mathcal{B}_{di}-\textbf L_i \bar{\mathcal D}_{di})+\bar{\mathcal D}_{di})\notag
\end{align}
There are two parts in this equation (\ref{7}): The first part depends on the gain matrix $\Delta L_i$ and the second part regarding the gain matrix $\textbf L_i$, which is calculated by \textit{Lemma \ref{l1}}. 

The transfer matrix $\bar{\mathcal C}_{i} (sI-\mathcal{A}_i+\textbf L_i \bar{\mathcal C}_{i})^{-1}(\mathcal{B}_{di}-\textbf L_i \bar{\mathcal D}_{di}) \in \mathbb{RH}_2$. We have
\begin{align*}
    \|\bar{\mathcal C}_{i} (sI-\mathcal{A}_i+\textbf L_i \bar{\mathcal C}_{i})^{-1}(\mathcal{B}_{di}-\textbf L_i \bar{\mathcal D}_{di}) \|_2^2 =trace (\bar{\mathcal C}_{i} Y_i \bar{\mathcal C}_{i}^T)
\end{align*}
is achieved by 
\begin{align*}
    \textbf L_i=Y_i\bar{\mathcal C}_{i}^T+\mathcal{B}_{di} \bar{\mathcal D}_{di}^T
\end{align*}
and matrix $Y_i \geq 0$ is solution of the Riccati equation
\begin{align*}
&(\mathcal{A}_i-\mathcal{B}_{di}\bar{\mathcal D}_{di}^T\bar{\mathcal C}_{i})Y_i+Y_i(\mathcal{A}_i-\mathcal{B}_{di}\bar{\mathcal D}_{di}^T\bar{\mathcal C}_{i})^T-Y_i\bar{\mathcal C}_{i}^T\bar{\mathcal C}_{i}Y_i\\
&+\mathcal{B}_{di}\mathcal{B}_{di}^T-\mathcal{B}_{di}\bar{\mathcal D}_{di}^T\bar{\mathcal D}_{di}\mathcal{B}_{di}^T =0
\end{align*}

When the disturbances (unknown inputs) are assumed to have known fixed spectral densities, the transfer matrix of disturbance $T_{r_{i}d_{Ni}}\in \mathbb{RH}_2^{\xi_{fi} \times \xi_{di}}$ can be calculated by $\mathcal H_2$ norm, so that $  \|T_{r_{i}d_{Ni}}\|_2^2 < \gamma_1^2$ which corresponds with the performance \textit{I} in (\ref{a}).

Note that $T_i=\bar{\mathcal C}_{i} (sI-\mathcal{A}_i+\textbf L_i \bar{\mathcal C}_{i})^{-1}(\mathcal{B}_{di}-\textbf L_i \bar{\mathcal D}_{di})+\bar{\mathcal D}_{di} \in \mathbb{RH}_{\infty}^{\xi_{fi} \times \xi_{di}}$ and $T_i\bar{\mathcal C}_{i}(sI-\mathcal{A}_i+\textbf L_i \bar{\mathcal C}_{i}+\Delta L_i \bar{\mathcal C}_{i})^{-1}(-\Delta L_i) \in \mathbb{RH}_{\infty}^{\xi_{fi} \times \xi_{di}}$. Thus, the norm of matrix transfer function of $T_{r_{i}d_{Ni}}$ finally has
\begin{align}\label{8}
\|T_{r_{i}d_{Ni}}\|_2^2 &= \|\bar{\mathcal C}_{i} (sI-\mathcal{A}_i+\textbf L_i \bar{\mathcal C}_{i})^{-1}(\mathcal{B}_{di}-\textbf L_i \bar{\mathcal D}_{di}) +\bar{\mathcal D}_{di}\|_2^2\notag\\
&+ \|\bar{\mathcal C}_{i}(sI-\mathcal{A}_i+\textbf L_i \bar{\mathcal C}_{i}+\Delta L_i \bar{\mathcal C}_{i})^{-1}(\Delta L_i) \|_2^2 
\end{align}
and determine the $\mathcal H_2$ norm of $\|\bar{\mathcal C}_{i} (sI-\mathcal{A}_i+\textbf L_i \bar{\mathcal C}_{i})^{-1}(\mathcal{B}_{di}-\textbf L_i \bar{\mathcal D}_{di}) \|_2^2$ with assumption $T_i \in \mathbb{RH}_2^{\xi_{fi} \times \xi_{di}}$,
\begin{align*}
&\|\bar{\mathcal C}_{i} (sI-\mathcal{A}_i+\textbf L_i \bar{\mathcal C}_{i})^{-1}(\mathcal{B}_{di}-\textbf L_i \bar{\mathcal D}_{di}) \|_2^2 \\
&\hspace{4cm}=trace (\bar{\mathcal C}_{i} Y_i \bar{\mathcal C}_{i}^T)\\
&\|\bar{\mathcal C}_{i}(sI-\mathcal{A}_i+\textbf L_i \bar{\mathcal C}_{i}+\Delta L_i \bar{\mathcal C}_{i})^{-1}(\Delta L_i) \|_2^2\\
&\hspace{4cm}=trace (\Delta L_i^T P_i\Delta L_i)
\end{align*}
According to (\ref{8}), the constraint $  \|T_{r_{i}d_{Ni}}\|_2^2 < \gamma_1^2$ becomes
\begin{align}\label{11}
trace(\Delta L_i^T P_i\Delta L_i) < \gamma_1^2 -trace (\bar{\mathcal C}_{i} Y_i \bar{\mathcal C}_{i}^T)
\end{align} 
where $P_i=P_i^T > 0$ is solution of Lyapunov equation 
\begin{align}\label{14a}
&[\mathcal{A}_i -(\textbf L_i +\Delta L_i)\bar{\mathcal C}_{i}]P_i \notag\\
&+P_i[\mathcal{A}_i -(L_i +\Delta L_i)\bar{\mathcal C}_{i}]^T + \mathcal{C}^T_i\bar{\mathcal C}_{i} =0
\end{align}
Using the new variable $Q_i$ such that  $trace (\Delta L_i^T P_i\Delta L_i) < trace ({Q_i})$, then \eqref{11} becomes
\begin{align}\label{16d}
 trace ({Q_i}) \leq \gamma_1^2 -trace (\bar{\mathcal C}_{i} Y_i \bar{\mathcal C}_{i}^T)
 \end{align}
where a real symmetric matrix $Q_i>0$  satisfied 
\begin{align}\label{16}
{Q_i}-\Delta L_i^T P_i\Delta L_i \ge 0
\end{align}
which implies that (\ref{11}) is satisfied. Then, by using Shur complement formula, if there exits a real symmetric matrix $P_i >0$ satisfying (\ref{11}), then (\ref{16}) is equivalent to
\begin{align}\label{15a}
\begin{bmatrix}
    	{{Q_{i}}}&{{\Delta L_i^{T}P_i}}\\
	{{P_i\Delta L_i}}&{{P_{i}}} 
\end{bmatrix} \geq 0
\end{align}

From the equation (\ref{14a}). Let's set $\mathcal{A}_i-({L}_i +\Delta L_i) \bar{\mathcal C}_{i} =\mathbf{A}_i -\Delta L_i \bar{\mathcal C}_{i}$, where $\mathbf{A}_i=\mathcal{A}_i - \textbf L_i \bar{\mathcal {C}}_i$. Using the performance \textit{III} in (\ref{a}),  $\mathbf{A}_i -\Delta L_i \bar{\mathcal C}_{i}$  is stable if and only if
\begin{align}\label{15d}
(\mathbf{A}_i -\Delta L_i\bar{\mathcal C}_{i})P_i +P_i(\mathbf{A}_i -\Delta L_i\bar{\mathcal C}_{i})^T + \mathcal{C}^T_i\bar{\mathcal C}_{i} < 0
\end{align}

The second step, the transfer matrix of faults with assumption that $T_{r_if_{Ni}}=\bar{\mathcal C}_{i} (sI-\mathcal{A}_i+(\textbf L_i+\Delta L_i) \bar{\mathcal C}_{i})^{-1}(\mathcal{B}_{fi}-(\textbf L_i+\Delta L_i) \bar{\mathcal D}_{fi})+\bar{\mathcal D}_{fi} \in \mathbb{RH}_{\infty}^{\xi_{fi} \times \xi_{fi}}$ is injective. Using the $\|T_{r_if_{Ni}}\|_{-} > \gamma_2$ in \textit{Lemma \ref{l2}}, which corresponds with the performance \textit{II} in (\ref{a}). 
Note that $\mathbf{A}_i=\mathcal{A}_i - \textbf L_i \bar{\mathcal {C}}_i, \mathbf{B}_{fi}=\mathcal{B}_{fi} - \textbf L_i \bar{\mathcal {C}}_i$, we have
\begin{align}\label{16a}
 &X_i(\mathbf{A}_i-\Delta L_i\bar{\mathcal C}_{i})+(\mathbf{A}_i-\Delta L_i\bar{\mathcal C}_{i})^TX_i+\bar{\mathcal C}_{i}^T \bar{\mathcal C}_{i}\notag\\
 &+(X_i(\mathbf{B}_{fi}-\Delta L_i\bar{\mathcal D}_{fi})+\bar{\mathcal C}_{i}^T \bar{\mathcal D}_{fi})(\gamma_2^2I-\bar{\mathcal D}_{fi}^T\bar{\mathcal D}_{fi})^{-1}\notag\\
 &\times((\mathbf{B}_{fi}-\Delta L_i\bar{\mathcal D}_{fi})^TX_i+\bar{\mathcal D}_{fi}^T\bar{\mathcal C}_{i} )>0
\end{align}
Using the Shur complement, the (\ref{16a}) can be rewritten as
\begin{align}\label{16f}
\begin{bmatrix}
    {{\bar{\mathcal D}_{fi}^T \bar{\mathcal D}_{fi}-\gamma_2^2 I}}&{{{X}_i(\mathbf{B}_{fi}-\Delta L_i \bar{\mathcal D}_{fi})+\bar{\mathcal C}_{i}^T \bar{\mathcal D}_{fi}}}\\
	{{* }}&{{F_i}} 
\end{bmatrix}>0
\end{align}
where $F_i=X_i(\mathbf{A}_i-\Delta L_i\bar{\mathcal C}_{i})+(\mathbf{A}_i-\Delta L_i\bar{\mathcal C}_{i})^TX_i+\bar{\mathcal C}_{i}^T \bar{\mathcal C}_{i}$.

Hence, using the (\ref{16d}), (\ref{15a}), (\ref{15d}) and  (\ref{16f}), the following constrained optimization problem: Find the matrices $\Delta L_i, $ such that
\begin{align}\label{16b}
&\hspace{0.5cm}\mathop {maximize}\limits_{\Delta L_i} \beta_2 \gamma_2-\beta_1 \gamma_1\\
&\text{subject to}\notag\\
&\hspace{0.5cm}(\ref{16d}), (\ref{15a}), (\ref{15d}) \;\text{and}\;  (\ref{16f})\notag
\end{align}
However, these matrix inequalities are nonlinear in terms of $\Delta L_i, P_i, Q_i, $ and $X_i$. To overcome this problem, instead of the search over matrices $\Delta L_i, P_i, Q_i, $ and $X_i$, a larger space of matrices will be found by introducing the change of variable $\Delta L_i = P^{-1}_i  N_i$ and assuming $X_i=P_i$. Therefore, the non-convex optimization problem (\ref{16b}) can be rewritten in another non-convex optimization form: Search $N_i$  such as \eqref{10}--\eqref{5b}.  This is end of the proof. 
\end{proof}

The \eqref{10}--\eqref{5b} are non-convex optimization problem.  Instead of finding $\gamma_1, \gamma_2$,  the optimization problem can be solved with $\gamma_1^2$ and $\gamma_2^2$. Let set $\alpha_1=\gamma_1^2$ and $\alpha_2=\gamma_2^2$, the non-convex optimization problem \eqref{10}--\eqref{5b} become convex optimization form.
\subsection{Residual evaluation and threshold setting}
Following the generation of the residuals $r_i(t) \in \mathbb{R}^{\xi_{fi}}, \forall i \in V$, the next step in the distributed FDI methodology is to figure out the threshold function $J_{{th}_i}$ and the evaluation function $J_{{r}_i}(t)$. In this work, the threshold function $J_{{r}_i}$ and the residual evaluation $J_{{th}_i}$ are selected as
\begin{align} \label{17}
J_{th_i}&=\sup_{f_{Ni}=0, u_{Ni}, d_{Ni}}\| J_{r_i}(t)\|_{\infty}\\
J_{r_{ij_k}}(t) &= \|r_{ij_k}(t)\|_{2,L}\nonumber \\
&=\left[1/L \sum_{s=t-L}^{t} r^T_{ij_k}(s)r_{ij_k}(s)\right]^{1/2}\label{17a}
\end{align}
where $r_{ij_k}(t)$ and $J_{r_{ij_k}}(t)$ are the $k^{th}, k= 1,...,\mu_{yi}$ elements of $j^{th}$ elements (neighbor agents) of the residual signal $r_i(t)$ and evaluation function $J_{r_i}(t)$, and $L$ is the evaluation finite time window. The occurrence of a fault can then be detected by using the following decision logic: If $J_{r_i}(t) > J_{th_i}$, then $f_i(t) \ne 0$ or $f_j(t) \ne 0, j \in N_i$.
\subsection{Distributed fault isolation}
The final step, to determine the faulty agent in the team since a fault is detected. Based on the  fault isolation techniques \cite{Davoodi2016}, the distributed strategy is proposed derived from flags, which are generated corresponding to the residual signal of the team agents. The so-called flags $[\epsilon_{ij}]_k, j \in V, j \in \{i,N_i\}$ in which $[\epsilon_{ij}]_k \in \mathbb R^{\mu_{yi}}$ is row vector and the $ij_k^{th}$ flag is 1 if $J_{r_{ij_k}}(t) > J_{th_i}$ and 0 otherwise. Consequently, it is assume that each agent constructs the defined fault pattern 
\begin{align}\label{29}
    \Upsilon_i=\left\{[\epsilon_{ij}]_k: j \in \{i,N_i\}, i \in V\right\}
\end{align}
Finally, the faulty agent can be isolated according to the following algorithm 1.
\begin{algorithm}
\SetAlgoLined   
\caption{A distributed fault isolation procedure}
1. Calculate $J_{r_{ij_k}}(t)$, $J_{th_i}$ in \eqref{17} and \eqref{17a}.
		
2. Determine the flags and fault pattern $\Upsilon_i$ in \eqref{29}.
	    
\If{all of elements of $\Upsilon_i=1$ (i.e., $[\epsilon_{ij}]_k =1$)}{
The $i^{th}$ agent is faulty
}
\If{only one element of $\Upsilon_i=1$}{
$j^{th}$ neighbor of agent $i^{th}$ is faulty
}
\If{$\Upsilon_{q_1}=\Upsilon_{q_n}=1, q_n \le i$ and the first element of $\Upsilon_{p_1},..,\Upsilon_{p_m}= 1 < q_n \ne p_m \le i$  corresponding to flags $[\epsilon_{ij}]_k$} {the fault will occur on the neighbor of agent $j$: It is agent $i$
}
\label{Al1}
\end{algorithm}
\section{Simulation Results}
\begin{figure}
	\centering
	\includegraphics[height=3.6cm]{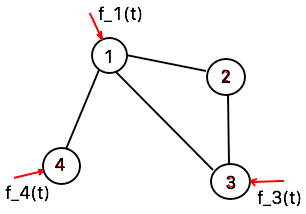}
    \caption{The topology of an MASs}\label{fig:1}
\end{figure}

In this section, a heterogeneous MASs with two first-order agent 1 and agent 2 as well as one second-order agent 4 and one fourth-order agent 3 are considered. Their topology graph is shown in Fig. \ref{fig:1}. The following data associated with continuous-time model (\ref{1}) is considered.
\begin{align*}
A_{1}&=\begin{bmatrix}
{{-5}}&{{0.05}}\\
{{0}}&{{-13}} 
\end{bmatrix}, B_{1}=\begin{bmatrix}
{{1}}\\
{{0}}
\end{bmatrix}, C_{1}=\begin{bmatrix}
{{3}}&{{0}}\\
{{0.01}}&{{0.1}}
\end{bmatrix},\\
B_{f1}&=\begin{bmatrix}
{{0.3}}\\
{{0.1}}
\end{bmatrix}, B_{d1}=\begin{bmatrix}
{{0.2}}\\
{{0.1}}
\end{bmatrix}, \\
D_{f1}&=\begin{bmatrix}
{{0.45}}\\
{{0.2}}
\end{bmatrix}, D_{d1}=\begin{bmatrix}
{{0.27}}\\
{{0.2}}
\end{bmatrix}\\
A_{2}&=\begin{bmatrix}
{{-2}}&{{0}}\\
{{0}}&{{-10}} 
\end{bmatrix}, B_{2}=\begin{bmatrix}
{{1}}\\
{{0.1}}
\end{bmatrix}, C_{2}=\begin{bmatrix}
{{1}}&{{0}}\\
{{0}}&{{0.1}}
\end{bmatrix},\\
B_{f2}&=\begin{bmatrix}
{{1}}\\
{{0.1}}
\end{bmatrix}, B_{d2}=\begin{bmatrix}
{{0.3}}\\
{{0.1}}
\end{bmatrix},\\
D_{f2}&=\begin{bmatrix}
{{0.15}}\\
{{0.3}}
\end{bmatrix}, D_{d2}=\begin{bmatrix}
{{0.47}}\\
{{0.1}}
\end{bmatrix},\\
A_{3}&=\begin{bmatrix}
{{-1}}&{{0}}&{{0.05}}&{{0}}\\
{{0}}&{{-3.6}}&{{0}}&{{0.1}}\\
{{0}}&{{0}}&{{-2}}&{{0.1}}\\
{{-2}}&{{-10}}&{{-13}}&{{-9}}
\end{bmatrix}, B_{3}=\begin{bmatrix}
{{1}}\\
{{0}}\\
{{1}}\\
{{0}}
\end{bmatrix}, \\
B_{f3}&=\begin{bmatrix}
{{0.1}}\\
{{0.2}}\\
{{0}}\\
{{0}}
\end{bmatrix}, B_{d3}=\begin{bmatrix}
{{0.2}}\\
{{0.1}}\\
{{0}}\\
{{0}}
\end{bmatrix},D_{d3}=\begin{bmatrix}
{{0.35}}\\
{{0.2}}
\end{bmatrix},\\
C_{3}&=\begin{bmatrix}
{{0.7}}&{{0.5}}&{{0}}&{{0}}\\
{{0.5}}&{{0.1}}&{{0}}&{{0.1}}
\end{bmatrix}, D_{f3}=\begin{bmatrix}
{{0.17}}\\
{{1}}
\end{bmatrix}\\
A_{4}&=\begin{bmatrix}
{{-1}}&{{0}}&{{0.05}}\\
{{0}}&{{-3.6}}&{{0}}\\
{{-2}}&{{-10}}&{{-13}}
\end{bmatrix}, B_{4}=\begin{bmatrix}
{{1}}\\
{{0}}\\
{{1}}
\end{bmatrix}, \\
B_{f4}&=\begin{bmatrix}
{{0.11}}\\
{{0.2}}\\
{{0}}
\end{bmatrix}, B_{d4}=\begin{bmatrix}
{{0.1}}\\
{{0.2}}\\
{{0}}
\end{bmatrix},D_{d4}=\begin{bmatrix}
{{0.15}}\\
{{0.3}}
\end{bmatrix},\\ 
C_{4}&=\begin{bmatrix}
{{0.7}}&{{0.5}}&{{0.1}}\\
{{0.5}}&{{0.1}}&{{0.1}}
\end{bmatrix}, D_{f4}=\begin{bmatrix}
{{0.27}}\\
{{1}}
\end{bmatrix}
\end{align*}
the disturbances $d_1(t), d_2(t), d_3(t)$ and $d_4(t)$ are band-limited white noise with powers 0.001. The input signals $u_1(t), u_2(t), u_3(t)$ and $u_4(t)$ are taken as step inputs with amplitudes of 0.1, 0.5, 1 and -1, respectively. The initial states of agents 1,2,3 and 4 are $[0.4; \;0.4]$, $[0.2; \;0.2]$, $[0.4; -0.3; \;0.4; \;0.2]$ and $[-0.3; \;0.3;\;0.3]$, respectively. The initial states of observers of all agents are set be zero. 
\begin{figure}
    \centering
    \includegraphics[height=3.5cm]{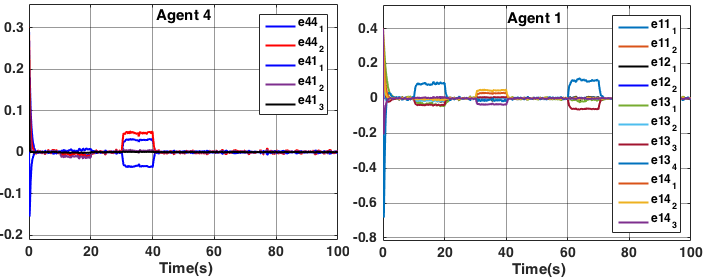}
    \caption{The error of estimation of agent 4 and agent 1}
    \label{fig:12}
\end{figure}
To demonstrate the performance of our proposed methodologies through considering different types of faults. $d_1(t), d_2(t), d_3(t)$ and $d_4(t)$ are assumed to occur in the agent 1, 2, 3 and 4, respectively. The fault signals in agent 1 ($f_1(t)$), agent 4 ($f_4(t)$) and agent 3 ($f_3(t)$) are simulated as a rectangular pulsed signal with an amplitude of -0.25, 0.5 and 0.15 that are connected during the time interval $[10 - 20]s$, $[30 - 40]s$ and $[60 - 70]s$, respectively. There is no faults in the agent 2 ($f_2(t)=0$). The estimation state error $e_{ijk}(t)=x_{Ni}(t) - \hat x_{Ni}(t)$ of agent 1 and agent 4 are represented in Fig. \ref{fig:12}. It shown that each agent estimates its own states and the states of its nearest neighbors in the presence of the disturbance $d_{Ni}(t)$, faults $f_{Ni}(t)$ and the control inputs $u_{Ni}(t)$. 
\begin{figure*}
  \centering
	\begin{minipage}{0.8\textwidth}
\includegraphics[height=9.1cm]{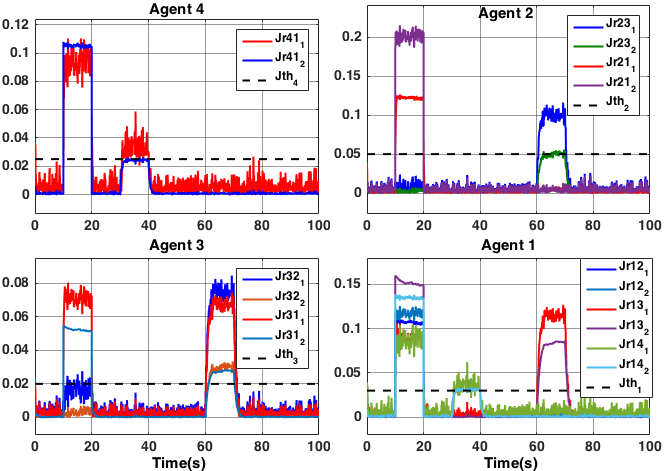}
\caption{Actuator Faults: Residual signals of agent 1,2,3 and agent 4}
\label{fig:8}
\end{minipage}\\
\vspace{1cm}
	\begin{minipage}{0.8\textwidth}
\includegraphics[height=9.1cm]{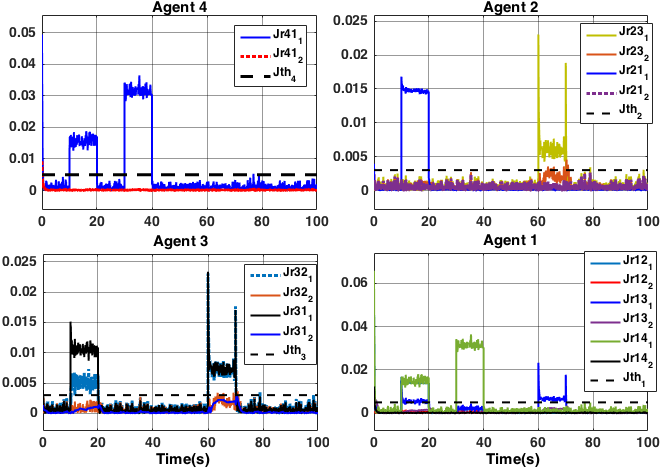}
\caption{Sensor Faults: Residual signals of agent 1,2,3 and agent 4}
\label{fig:8a}
\end{minipage}
\end{figure*}

To illustrate the effectiveness of our proposed method solving distributed FDI problem for system with parameter matrices given above. The relative mode system should be constructed for agent 1, 2, 3 and 4, respectively. Faults will take into account in two scenarios: actuator faults $\mathcal{B}_{fi} =\mathcal B_{ui}, \bar{\mathcal D}_{fi} =\bar{\mathcal D}_{di}=0, \; i=1,2,3,4$ and sensor faults $\mathcal{B}_{fi}=0, \bar{\mathcal D}_{fi} =I, \; i=1,2,3,4$, respectively. 
\subsection{Scenario 1: Actuator Faults}
In the following, we illustrate the performance of our proposed methodology by considering actuator fault. The thresholds are obtained corresponding to the threshold functions of the agents in our method. In \textit{Theorem \ref{thm1}}: $J_{th1}=0.03, J_{th2}=0.05, J_{th3}=0.02,$ and $J_{th4}=0.025,$. Moreover, the residual evaluation functions $J_{r_1}(t), J_{r_2}(t), J_{r_3}(t),$ and $J_{r_4}(t)$ are shown in Fig. \ref{fig:8}. It is clearly realized that each fault can be detected from the other faults and the disturbances. The next step, fault can be isolated by using the Algorithm 1 for each method of our proposed methodologies.

For the fault $f_4(t)$, the fault pattern $\Upsilon_i$ are obtained as $\Upsilon_1= \left\{ \begin{bmatrix}
{{0}}\\
{{0}}
\end{bmatrix}, \begin{bmatrix}
{{0}}\\
{{0}}
\end{bmatrix}, \begin{bmatrix}
{{1}}\\
{{1}}
\end{bmatrix} \right\}, \Upsilon_2= \left\{ \begin{bmatrix}
{{0}}\\
{{0}}
\end{bmatrix},  \begin{bmatrix}
{{0}}\\
{{0}}
\end{bmatrix}\right\}, 
\Upsilon_3= \left\{ \begin{bmatrix}
{{0}}\\
{{0}}
\end{bmatrix}, \begin{bmatrix}
{{0}}\\
{{0}}
\end{bmatrix} \right\},
\Upsilon_4= \begin{bmatrix}
{{1}}\\
{{1}}
\end{bmatrix}$. All of elements of $\Upsilon_4=1$ that means the fourth agent is thus faulty, and the agent 1 is the nearest neighbor to the agent 4 based on observing the fault pattern $\Upsilon_1$.  Moreover, the fault patterns for the fault signal $f_1(t)$ are obtained as $\Upsilon_1= \left\{ \begin{bmatrix}
{{1}}\\
{{1}}
\end{bmatrix},  \begin{bmatrix}
{{1}}\\
{{1}}
\end{bmatrix},\begin{bmatrix}
{{1}}\\
{{1}}
\end{bmatrix} \right\}
, \Upsilon_2=\begin{bmatrix}
{{1}}\\
{{1}}
\end{bmatrix},  \begin{bmatrix}
{{0}}\\
{{0}}
\end{bmatrix}, 
\Upsilon_3= \left \{ \begin{bmatrix}
{{1}}\\
{{1}}
\end{bmatrix},  \begin{bmatrix}
{{0}}\\
{{0}}
\end{bmatrix} \right \},
\Upsilon_4= \left \{\begin{bmatrix}
{{1}}\\
{{1}}
\end{bmatrix} \right \}$. It recognize that there are two fault patterns $\Upsilon_1, \Upsilon_4$ which have all of elements equal to one. In addition, the first element of fault patterns $\Upsilon_2= \Upsilon_3=1$ corresponding to flags $[r_{21}], [r_{31}]$ this mean that the fault will occur on the neighbor of agent 2 and agent 3, respectively. Consequently, the agent 1 is faulty and agent 2, 3, 4 are the nearest neighbors of agent 1, respectively. Besides, the fault patterns for the fault signal $f_3(t)$ are determined such that $\Upsilon_1= \left\{ \begin{bmatrix}
{{0}}\\
{{0}}
\end{bmatrix},  \begin{bmatrix}
{{1}}\\
{{1}}
\end{bmatrix}, \begin{bmatrix}
{{0}}\\
{{0}}
\end{bmatrix}\right\}
, \Upsilon_2= \begin{bmatrix}
{{0}}\\
{{0}}
\end{bmatrix},  \begin{bmatrix}
{{1}}\\
{{1}}
\end{bmatrix}, 
\Upsilon_3= \left\{ \begin{bmatrix}
{{1}}\\
{{1}}
\end{bmatrix}, \begin{bmatrix}
{{1}}\\
{{1}}
\end{bmatrix} \right \},
\Upsilon_4= \begin{bmatrix}
{{0}}\\
{{0}}
\end{bmatrix}$ and it conclude that the agent 3 is thus faulty because of all of elements of $\Upsilon_3=1$. Besides, the agent 1 and agent 2 are the nearest neighbors to the agent 4, respectively.
\subsection{Scenario 2: Sensor Faults}
In the subsection, the sensor fault will be investigated based on our proposed approach (which is can not implemented using the unknown input observer as \cite{Liu2016}, \cite{Quan2016}, \cite{Shames2011}). Similarly as mention above, the thresholds are obtained corresponding to the threshold functions of the agents in our method $J_{th1}=0.005, J_{th2}=0.0025, J_{th3}=0.0025,$ and $J_{th4}=0.005$. Moreover, the residual evaluation functions $J_{r_1}(t), J_{r_2}(t), J_{r_3}(t)$ and $J_{r_4}(t)$ are shown in Fig. \ref{fig:8a}. The analyses are the same in Actuator Faults.
\section{Conclusions}
In this work, robust distributed observer-based on Fault Detection and Isolation for a network of heterogeneous MASs is investigated. By using the proposed methodology not only each agent's faults (Actuator faults, and Sensor faults) but also faults of the agent's nearest neighbors can also be detected and isolated. Furthermore, the relative model system is utilized by a vector of relative output measurements. Each agent is thus able to estimate not only its own states but also states of its nearest neighbors and the dimension of observer at each node is reduced. Based on two performance indexes $\mathcal H_2$ and $\mathcal H_{-}$, such as a set of linear matrix inequality conditions, sufficient conditions for solvability of the optimization problem were obtained. 

\bibliographystyle{IEEEtran}
\bibliography{ref.bib}%
\end{document}